\documentclass[trackchanges,twocolumn]{aastex7}
\graphicspath{{./}{figures/}}

\usepackage[acronym,nohypertypes={acronym}]{glossaries} 
\newacronym{LSST}{LSST}{Legacy Survey of Space and Time}
\newacronym{JWST}{JWST}{James Webb Space Telescope}
\newacronym{HST}{HST}{Hubble Space Telescope}
\newacronym{DM}{DM}{dark matter}
\newacronym{GZ}{GZ}{Galaxy Zoo}
\newacronym{SNR}{SNR}{signal-to-noise ratio}
\newacronym{TW}{TW}{Tremaine-Weinberg}
\newacronym{IFU}{IFU}{integral field unit}
\newacronym{MaNGA}{MaNGA}{Mapping Nearby Galaxies at Apache Point Observatory}
\newacronym{JAM}{JAM}{Jeans anisotropic modeling}
\newacronym{MGE}{MGE}{Multi-Gaussian Expansion}
\newacronym{DynPop}{DynPop}{Dynamics and stellar Population }
\newacronym{SPS}{SPS}{stellar population synthesis}
\newacronym{IMF}{IMF}{initial mass function}
\newacronym{AGN}{AGN}{active galactic nucleus}
\newacronym{CALIFA}{CALIFA}{Calar Alto Legacy Integral Field Area}
\newacronym{ALFALFA}{ALFALFA}{Arecibo Legacy Fast ALFA}
\newacronym{M/L}{M/L}{mass-to-light ratio}

\newcommand{\glswithcite}[2]{%
    \ifglsused{#1}{\gls{#1} (#2)}{\glsentrylong{#1} (\glsentryshort{#1}; #2)\glsunset{#1}}
}

\newcommand{\glsplwithcite}[2]{%
    \ifglsused{#1}{\glspl{#1} (#2)}{\glsentrylongpl{#1} (\glsentryshortpl{#1}; #2)\glsunset{#1}}
}

\newcommand{\todo}[1]{{#1}}

\shorttitle{The Role of Baryonic and Dark Matter in Bar Kinematics}
\shortauthors{G\'eron et al.}

\begin{document}


\title{The Role of Baryonic and Dark Matter in Bar Kinematics}

\author[orcid=0000-0002-6851-9613,gname=Tobias,sname=G\'eron]{Tobias G\'eron}
\affiliation{Dunlap Institute for Astronomy \& Astrophysics, University of Toronto, 50 St. George Street, Toronto, ON M5S 3H4, Canada}
\email[show]{tobias.geron@utoronto.ca} 

\author[0000-0002-2583-2669]{Kai Zhu}
\affiliation{Department of Astronomy, Westlake University, Hangzhou 310030, Zhejiang Province, People’s Republic of China}
\affiliation{Westlake Institute for Advanced Study, Hangzhou 310030, Zhejiang Province, People’s Republic of China}
\email{}

\author[0000-0001-9520-7765]{Steph Campbell}
\affiliation{School of Mathematics, Statistics and Physics, Newcastle University, Newcastle upon Tyne, NE1 7RU, UK}
\email{}

\author[0000-0001-5578-359X]{Chris Lintott}
\affiliation{Oxford Astrophysics, Department of Physics, University of Oxford, Denys Wilkinson Building, Keble Road, Oxford, OX1 3RH, UK}
\email{}

\author[0000-0001-6417-7196]{R. J. Smethurst}
\affiliation{Oxford Astrophysics, Department of Physics, University of Oxford, Denys Wilkinson Building, Keble Road, Oxford, OX1 3RH, UK}
\email{}

\author[0000-0002-1584-2281]{Behzad Tahmasebzadeh}
\affiliation{Department of Astrophysics and Planetary Science, Villanova University, 800 East Lancaster Ave., Villanova, PA 18085, USA}
\email{}



\begin{abstract}
Simulations predict that bars in galaxies should slow down over time. This is often attributed to the exchange of angular momentum between the bar and other regions of the galaxy, such as the outer disc and dark matter halo, which implies that galaxies with a more massive halo or disc should be able to slow down the bar more efficiently. However, observational evidence for this process has been limited. In this work, we provide observational support for the slowing down of bars as predicted by simulations. We combine bar kinematics measurements obtained with the Tremaine-Weinberg method and host galaxy mass estimates derived from Jeans anisotropic models for a sample of \todo{30} galaxies from the MaNGA survey. We find a statistically significant anti-correlation (\todo{$>4\sigma$}) between the bar pattern speed and both the stellar and total dynamical mass, which suggests that the slowest bars reside in the most massive galaxies. However, while the slope of the best-fit line between the pattern speed and dark matter mass is negative, it is not statistically significant (\todo{$2.43\sigma$}). We also find that bars with lower pattern speeds have more extended NFW dark matter profiles with lower central densities. Additionally, we find statistically significant correlations (\todo{$>3\sigma$}) between the corotation radius and the stellar mass, dark matter mass, and total dynamical mass. Finally, we find no significant correlations that involve the dark matter fraction or $\mathcal{R}$, likely due to the inherent challenges associated with measuring these specific parameters accurately. 
\end{abstract}



\section{Introduction}
\label{sec:introduction}




Many disc galaxies contain a central elongated stellar structure, called a bar. These bars are very common; around 43\%-52\% of disc galaxies have a bar when studied with optical wavelengths \citep{marinova_2007,barazza_2008,aguerri_2009,buta_2019,geron_2021}. When using infrared wavelengths instead, this fraction rises up to 59-73\% \citep{eskridge_2000,marinova_2007,menendez_delmestre_2007,sheth_2008}, while fractions of 23.6 - 29.4\% are found when only considering strongest and largest bars \citep{masters_2011, skibba_2012, cheung_2013}. The bar fraction remains 10 - 20\% at $z = 2$ \citep{guo_2024,leconte_2024,geron_2025}, suggesting that they are common structures even at higher redshifts.

Bars are also able to significantly affect the evolution of their host galaxy. For example, they can redistribute angular momentum throughout the galaxy \citep{lynden_bell_1972,sellwood_1981,athanassoula_2003,athanassoula_2013} and funnel gas from the outskirts of the galaxy to its center \citep{sorensen_1976,athanassoula_1992,davoust_2004,villa-vargas_2010,fragkoudi_2016,vera_2016,spinoso_2017,george_2019}. This inflow of gas can cause a central burst of star formation \citep{jogee_2005,sheth_2005,hunt_2008} and possibly trigger an \glswithcite{AGN}{\citealp{fanali_2015,galloway_2015,garland_2024}}. Bars are also more often found in red, massive, and gas-poor galaxies \citep{hoyle_2011,masters_2011, masters_2012,cheung_2013,vera_2016,cervantessodi_2017,kruk_2018,fraser_mckelvie_2020b}, which suggests that they are involved in the quenching process. 

These bars are dynamic structures that rotate with a specific pattern speed ($\Omega_{\rm b}$), also known as the rotational frequency of the bar. Another important parameter is $\mathcal{R}$, the dimensionless ratio of the corotation radius to the radius of the bar. This parameter is often used to classify bars into fast ($1.0 < \mathcal{R} < 1.4$) and slow bars ($\mathcal{R} > 1.4$; e.g., see \citealp{debattista_2000,rautiainen_2008,aguerri_2015}). The kinematics of the bar is often measured in observations with the \gls{TW} method \citep{tremaine_1984} on \gls{IFU} data \citep{aguerri_2015,cuomo_2019,guo_2019, garma_oehmichen_2020,cuomo_2021,garma_oehmichen_2022,geron_2023}. For example, \citet{geron_2023} used the \gls{TW} method on data from the \gls{MaNGA} survey \citep{bundy_2015} to obtain measurements of the kinematics of the bar for 225 galaxies.



Interestingly, simulations predict that the pattern speed does not stay constant. After the bar is formed, it slows down while it grows longer and stronger. This slowdown is attributed to dynamical friction and the aforementioned exchange of angular momentum between the bar and other structures in the galaxy, such as the outer disc and dark matter halo \citep{lynden_bell_1972,sellwood_1981,athanassoula_2003,sellwood_2008,athanassoula_2013}, which suggests that galaxies with a more massive dark matter halo or disc should be able to slow down the bar more efficiently. This slowdown of the bar depends on properties of the host galaxy: the stellar mass, the mass of the stellar bulge, the gas fraction, the spin of the dark matter halo, and the density of the dark matter halo \citep{kataria_2019, beane_2023, li_2023, li_2024, semczuk_2024}.


However, observational evidence for this slowdown has been limited. \citet{buttitta_2023} measured the bar kinematics and dark matter fraction for two lenticular galaxies: NGC 4264 and NGC 4277. They find that the galaxy with the slower bar has a higher dark matter fraction. \citet{tahmasebzadeh_2024} also measured the bar kinematics and dark matter fraction for one barred S0 galaxy, NGC 4371. They measure a high dark matter fraction ($f_{\rm DM} = 0.51\pm0.06$) and find that the bar is slow ($\mathcal{R} = 1.88\pm0.37$). Both of these studies are in agreement with the aforementioned predictions made by simulations. Although these initial results are exciting, the combined sample size of these two studies is still low ($n=3$), which makes making generalizations to the entire population of barred galaxies unreliable. Interestingly, \citet{guo_2019} compared $\mathcal{R}$ with the dark matter fraction for a sample of 53 barred galaxies, but they found no evidence for the slowdown of bars due to the dark matter halo. However, they acknowledge that this correlation might be obscured as the slowdown process involves additional variables (e.g., the initial bar pattern speed and the age of bar), which are hard to correct for in observations with current techniques.  



In this work, we will look for evidence of the slowing down of bars. Previous observational studies either have very low sample sizes or do not find any statistically significant evidence for this process. We will expand on previous work by looking at multiple parameters that characterize bar kinematics (i.e., bar pattern speed, corotation radius, and $\mathcal{R}$) and different measures of the mass of the host galaxy (i.e., stellar mass, dark matter mass, total dynamical mass, and the dark matter fraction) for a large sample of galaxies.

The structure of the paper is as follows: the sample selection and methods are explained in Section \ref{sec:data_methods}. Section \ref{sec:results} shows our results, which are discussed in Section \ref{sec:discussion}. Finally, our conclusions are summarized in Section \ref{sec:conclusion}. Where necessary, we assumed a flat $\Lambda$CDM cosmology with cosmological parameters obtained from the Planck mission \citep{planck_2020}, implemented with \texttt{astropy} \citep{astropy_2013,astropy_2018,astropy_2022}. 

\section{Data and Methods}
\label{sec:data_methods}

This study combines estimates of bar kinematics measured with the \gls{TW} method with dynamical models from previous catalogs \citep{geron_2023, zhu_2023}. The \gls{TW} method and bar kinematics are explained in greater detail in Section \ref{sec:tw}, while the dynamical modeling is explained in \ref{sec:jam}. Finally, the sample selection is explained in \ref{sec:sample_selection}.

\subsection{Measuring Bar Kinematics with the Tremaine-Weinberg Method}
\label{sec:tw}

The kinematics of a bar is often described with three parameters: the bar pattern speed, corotation radius, and $\mathcal{R}$. In this work, we use the measurements of the bar kinematics performed by \citet{geron_2023} on data from the \gls{MaNGA} survey \citep{bundy_2015} using the \gls{TW} method. The measurements were made with the \texttt{Tremaine\_Weinberg} package\footnote{\url{https://github.com/tobiasgeron/Tremaine_Weinberg}}. Their sample includes pattern speeds, corotation radii, and $\mathcal{R}$ for 225 strongly and weakly barred galaxies identified with Galaxy Zoo DESI \citep{walmsley_2023}. It is the largest catalog of bar kinematics measurements to date and has been used in other work (e.g., see \citealp{geron_2024, mcclure_2025, pearlstein_2025, puczek_2025}).

The \gls{TW} method is a model-independent technique to measure the bar pattern speed of barred galaxies \citep{tremaine_1984}. It was initially developed for long-slit spectroscopy, but has been adapted to work on data from \glsplwithcite{IFU}{e.g., see \citealt{aguerri_2015,cuomo_2019,guo_2019,garma_oehmichen_2020,garma_oehmichen_2022}}. The \gls{TW} method assumes that there is a well-defined pattern speed and that the tracer used (i.e., stars or gas) satisfies the continuity equation. It can be expressed as:

\begin{equation}
    \Omega_{\rm b} \sin{\left( i \right)} = \frac{\left<V\right>}{\left<X\right>} \;,
\end{equation}

where $\Omega_{\rm b}$ is the bar pattern speed, $i$ is the inclination of the galaxy, $\left<V\right>$ is the kinematic integral, and $\left<X\right>$ is the photometric integral. These last two integrals are defined as: 

\begin{equation}
  \left<V\right> = \frac{\int_{\rm -\infty}^{\rm +\infty} V_{\rm LOS} (X,Y) \Sigma (X,Y) \text{d}\Sigma}{\int_{\rm -\infty}^{\rm +\infty} \Sigma (X,Y) \text{d}\Sigma} \; ,
  \label{eq:tw_V_eqs}
\end{equation}

\begin{equation}
  \left<X\right> = \frac{\int_{\rm -\infty}^{\rm +\infty} X \Sigma (X,Y) \text{d}\Sigma}{\int_{\rm -\infty}^{\rm +\infty} \Sigma (X,Y) \text{d}\Sigma} \;,
  \label{eq:tw_X_eqs}
\end{equation}

where $V_{\rm LOS}$ is the line of sight velocity of the galaxy and $\Sigma$ is the surface brightness of the galaxy. $\left<X\right>$ and $\left<V\right>$ are effectively the luminosity-weighted mean position and line of sight velocity, respectively. The coordinate system $\left(X,Y\right)$ is found in the sky plane with the origin centered on the center of the galaxy and the $X$-axis aligned with the major axis of the galaxy. The photometric and kinematic integrals are calculated for multiple pseudo-slits across the IFU, aligned parallel to the $X$-axis. 

The next important parameter is the corotation radius ($R_{\rm CR}$). This is the radius where the centrifugal and gravitational forces balance each other in the rest frame of the bar, which implies that the stars in the disc of the galaxy will have the same angular velocity as the bar at this radius. \citet{geron_2023} estimate the corotation radius by comparing the bar pattern speed to a model of the galaxy rotation curve, in this case the two-parameter arctan model described by \citet{courteau_1997}. Do note that the corotation radius is also often approximated as $R_{\rm CR} = V_{\rm c} / \Omega_{\rm b}$, where $V_{\rm c}$ is the circular velocity in the flat part of the rotation curve. This often approximates the corotation radius well, but assumes that the corotation radius lies in the region where the rotation curve has flattened, which can lead to biased estimates of $R_{\rm CR}$ when this is not the case.

The last parameter, $\mathcal{R}$, is a dimensionless parameter that is defined as the ratio of the corotation radius to the bar radius: $\mathcal{R} = R_{\rm CR} / R_{\rm bar}$. This parameter is often used to classify bars into fast ($1.0 < \mathcal{R} < 1.4$) and slow bars ($\mathcal{R} > 1.4$) (e.g., see \citealp{debattista_2000,rautiainen_2008,aguerri_2015}). The bar lengths (i.e., twice the bar radius) in \citet{geron_2023} were measured manually on \textit{grz}-images obtained from the DESI Legacy Imaging Surveys \citep{dey_2019} using DS9 \citep{joye_2003}. Please refer to \citet{geron_2023} for more technical details on the implementation of the \gls{TW} method and a step-by-step example of how the bar pattern speed, corotation radius, and $\mathcal{R}$ are calculated. 


\subsection{Dynamical Modeling with JAM}
\label{sec:jam}


We use the measurements of stellar mass ($M_{\rm *, Re}$), dark matter mass ($M_{\rm DM, Re}$), total dynamical mass ($M_{\rm T, Re}$) and dark matter fraction ($f_{\rm DM, Re}$) obtained by the \gls{DynPop} collaboration \citep{zhu_2023}. All of these measurements are made within the effective radius of the galaxy ($R_e$). \citet{zhu_2023} obtain these parameters by creating dynamical models of galaxies using the \gls{JAM} method \citep{cappellari_2008, cappellari_2020}. 

The first step in creating these dynamical models involves obtaining a description of the tracer density distribution. This is done in \citet{zhu_2023} by applying the \gls{MGE} method on SDSS $r$-band images of the galaxies using \texttt{MgeFit}\footnote{\url{https://pypi.org/project/mgefit/}} \citep{emsellem_1994, cappellari_2002}. As the name implies, this  effectively parameterizes the surface brightness of a galaxy as a sum of multiple two-dimensional Gaussians. Interestingly, \citet{zhu_2023} note that the presence of strong bars can affect the quality of the \gls{MGE} fitting process. However, they take a number of precautions to deal with this. For example, they use the kinematic position angle during the \gls{MGE} fitting process specifically to avoid the effect that bars have on photometric position angles. Furthermore, they use the updated \texttt{MGE\_FIT\_SECTORS\_REGULARIZED} routine of the \gls{MGE} formalism (see \citealt{scott_2009, scott_2013}). This updated routine also serves to minimize the effect of bars on the model results by restricting the range of possible axial ratios for the different Gaussian components in the \gls{MGE} fit.

\citet{zhu_2023} then develop the actual \gls{JAM} dynamical models using \texttt{JamPy}\footnote{\url{https://pypi.org/project/jampy/}} \citep{cappellari_2020}. This method starts with the general Jeans equations in cylindrical coordinates ($R$, $z$, $\phi$;  \citealp{jeans_1922,binney_1987}):

\begin{equation}
    \frac{\nu\overline{v_R^2} - \nu\overline{v_\phi^2}}{R} + \frac{\partial(\nu\overline{v_R^2})}{\partial R} + \frac{\partial(\nu\overline{v_R v_z})}{\partial z} = -\nu \frac{\partial\Phi}{\partial R} \;,
    \label{eq:jam_1}
\end{equation}

\begin{equation}
    \frac{\nu\overline{v_R v_z}}{R} + \frac{\partial(\nu\overline{v_z^2})}{\partial z} + \frac{\partial(\nu\overline{v_R v_z})}{\partial R} = -\nu \frac{\partial\Phi}{\partial z} \;,
    \label{eq:jam_2}
\end{equation}

where $\nu$ is the number density of the tracer population and $\Phi$ is the gravitational potential. Additional assumptions are made on the orientation of the velocity ellipsoid (which is either assumed to be spherically or cylindrically aligned) and on the anisotropy of the system. These assumptions simplify Equations \ref{eq:jam_1} and \ref{eq:jam_2} and ensure a unique and solvable solution. For more details and the reduced forms of these equations, see \citet{zhu_2023}. 




Crucially for this work, the \gls{JAM} method is able to distinguish between the stellar mass distribution and dark matter mass distribution in the gravitational potential. The former is obtained from the aforementioned \gls{MGE} parameterization of the observed stellar surface brightness. The dark matter mass distribution is obtained by assuming the well-established spherical NFW dark matter halo profile described by \citet{navarro_1996}, which is parameterized as:

\begin{equation}
    \rho_{\rm DM} (r) = \rho_s \left(\frac{r}{r_s}\right)^{-1} \left(\frac{1}{2} + \frac{1}{2} \frac{r}{r_s}\right)^{-2} \;,
  \label{eq:nfw}
\end{equation}

where $\rho_s$ is the characteristic density and $r_s$ is the characteristic radius. Note that in these \gls{JAM} models, $r_s$ is not actually a free parameter. Instead, it is calculated using the stellar-to-halo mass relation \citep{moster_2013} and mass–concentration relation \citep{dutton_2014}. However, $\rho_s$ is a free parameter.

It is worth noting that the \gls{JAM} method assumes axial symmetry, which is not the case for strongly barred galaxies. However, to make sure that we only work with reliable models that were not affected by the presence of a bar, we added a number of quality thresholds (see Section \ref{sec:sample_selection}). Furthermore, \gls{JAM} models of barred galaxies have been used successfully in previous work (e.g., see \citealt{guo_2019, buttitta_2023}). The effect that strong bars have on \gls{JAM} models was studied in greater detail in the simulations of \citet{lablanche_2012}. They found that the \gls{M/L} can be overestimated if the inclination of the galaxy is low ($i \lesssim 30^{\circ}$). \citet{lablanche_2012} also found that the \gls{M/L} from \gls{JAM} is unbiased for barred galaxies when $\vert \textrm{PA}_{\rm bar} - {\textrm{PA}_{\rm disc}} \vert$ = $45^{\circ}$. However,  an over/underestimation of up to 15 per cent can appear when $\vert \textrm{PA}_{\rm bar} - {\textrm{PA}_{\rm disc}} \vert$ deviates from the ideal $45^{\circ}$ to either lower ($\sim18^{\circ}$) or higher ($\sim87^{\circ}$) values. However, in this work, we have already excluded face-on galaxies as well as galaxies where the bar is either parallel or perpendicular to the major axis of the disc. This is because the \gls{TW} method is not applicable to these kinds of configurations (see \citet{geron_2023} for more detail). The mean inclination of the galaxies used in this work is $\todo{48.9}^{\circ}$, while the mean  $\vert \textrm{PA}_{\rm bar} - {\textrm{PA}_{\rm disc}} \vert$ is $\todo{39.3}^{\circ}$. Finally, the results of this paper do not change if we explicitly limit the sample to galaxies with $i > 45^{\circ}$ and $30^{\circ} < \vert \textrm{PA}_{\rm bar} - {\textrm{PA}_{\rm disc}} \vert < 60^{\circ}$, despite the decrease in sample size. This suggests that the results presented in this work are not significantly influenced by any biases in the \gls{JAM} models introduced by the bars.

In \citet{zhu_2023}, \gls{JAM} models are fit to the observed line-of-sight velocities of galaxies from the \gls{MaNGA} survey \citep{bundy_2015} to obtain reliable estimates of model parameters, from which we can derive the measurements used in this work: $\log \left(M_{\rm *, Re}\right)$, $\log \left(M_{\rm DM, Re}\right)$, $\log \left(M_{\rm T, Re}\right)$, and $f_{\rm DM, Re}$. For more technical details on the implementation of the \gls{JAM} method, we refer the reader to \citet{zhu_2023}. 


\subsection{Sample Selection}
\label{sec:sample_selection}


\citet{zhu_2023} provide \gls{JAM} dynamical models for all 10,010 unique galaxies in the main \gls{MaNGA} survey. However, not all the galaxies in \gls{MaNGA} can be accurately described with these models due to low \gls{SNR} or highly disturbed kinematics. The quality of the \gls{JAM} model fit was assessed by \citet{zhu_2023} by visually inspecting the output. Each model was given a quality level between -1 and 3, where a higher number corresponds to a more reliable model. As per recommendation by \citet{zhu_2023}, we remove all galaxies with a quality level of -1 or 0. We also only include galaxies where $f_{\rm DM, Re} > 1\textrm{e}^{-5}$, to avoid using targets where the lower parameter boundary limit was reached during the fitting process. This reduced the sample size to \todo{4,867} unique galaxies. 

\begin{figure*}
  \centering
    \includegraphics[width=1.0\textwidth]{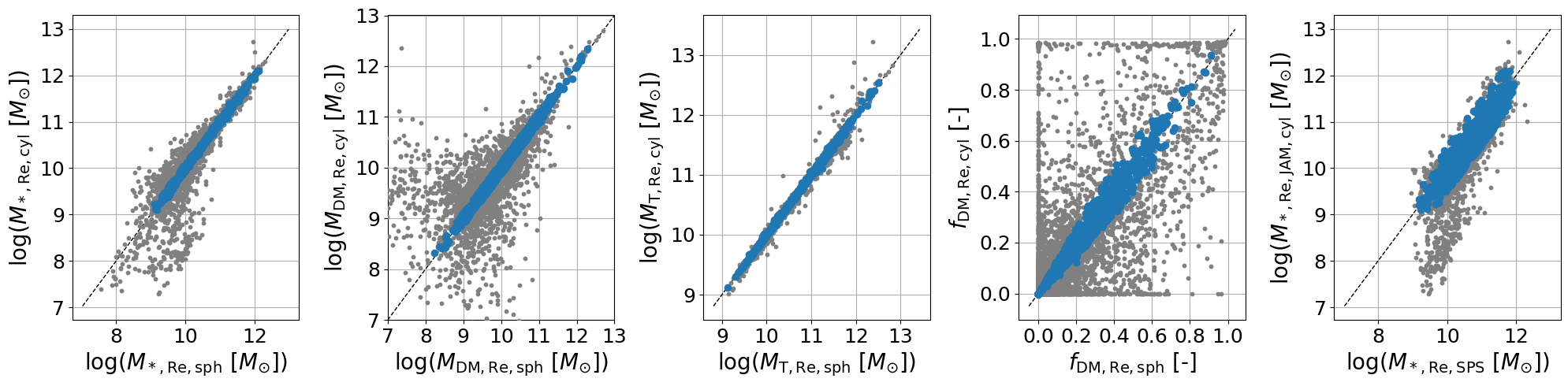}
    \caption{An overview of the consistency thresholds on the \gls{JAM} models used in this work. We compare the \gls{JAM} model outputs assuming a spherically-aligned velocity ellipsoid ($X_{\rm sph}$) and cylindrically-aligned velocity ellipsoid ($X_{\rm cyl}$) for the stellar mass (left panel), dark matter mass (second panel), total dynamical mass (third panel), and dark matter fraction (fourth panel). The rightmost panel shows a comparison between JAM-based and SPS-based stellar mass estimate. The galaxies that are kept in the sample after all consistency thresholds are shown in blue, while all the galaxies that do not pass these thresholds are shown in gray.}
    \label{fig:consistency}
\end{figure*}

We apply additional thresholds to ensure consistency of the model output. As noted in Section \ref{sec:jam}, the \gls{JAM} method assumes either a cylindrically or spherically aligned velocity ellipsoid, which results in two estimates for each model parameter (i.e., $f_{\rm DM, Re, cyl}$ and $f_{\rm DM, Re, sph}$ represent the dark matter fraction using the cylindrically and spherically aligned velocity ellipsoid, respectively). To guarantee consistency between the two models, as recommended by \citet{zhu_2023}, we continue to work only with galaxies that pass the following thresholds:

\begin{equation}
\begin{aligned}
    &\vert f_{\rm DM, Re, cyl} - f_{\rm DM, sph} \vert < 0.1\\
    &\vert \log \left(M_{\rm *, Re, cyl}\right) - \log \left(M_{\rm *, Re, sph}\right) \vert < 3\Delta\\
    &\vert \log \left(M_{\rm DM, Re, cyl}\right) - \log \left(M_{\rm DM, Re, sph}\right) \vert < 3\Delta\\
    &\vert \log \left(M_{\rm T, Re, cyl}\right) - \log \left(M_{\rm T, Re, sph}\right) \vert < 3\Delta
\end{aligned}
\end{equation}


where $\Delta$ is the observed scatter found by \citet{zhu_2023} for each parameter (see their Tables 2 and 3). These thresholds are visualized for our sample in the first four panels of Figure \ref{fig:consistency}. We continue to use the measurements that assume a cylindrically-aligned velocity ellipsoid for the rest of this work. However, the results presented in this paper do not change if we use the spherically-aligned velocity ellipsoid measurements instead. 




The \gls{JAM} method, and dynamical techniques in general, often struggle with estimating dark matter masses and fractions, due to the elusive nature of dark matter. As baryonic and dark matter both affect galaxy kinematics in identical ways, dynamical methods only directly model total dynamical masses. While it is possible to add assumptions on the shape of the dark matter halo and its contribution, some uncertainties still exist. This also becomes clear from looking at Figure \ref{fig:consistency}, where you can see that the scatter in the dark matter mass and the dark matter fraction estimates between the two \gls{JAM} models is larger than the scatter between the stellar and total dynamical masses. To break this degeneracy between the luminous and dark components, we continue to work only with galaxies where the stellar mass estimates from \gls{JAM} are consistent with the stellar mass estimates from another method. In this case, we turn to stellar mass estimates presented in \citet{lu_2023}, which is also part of the \gls{DynPop} collaboration. They calculated stellar masses using a \gls{SPS} approach. These stellar masses are estimated by fitting the galaxy spectra using predefined stellar populations and assuming a Salpeter \glswithcite{IMF}{\citealp{salpeter_1955}}. For more details on the \gls{SPS} measurements, see \citealp{lu_2023}. These \gls{SPS}-based stellar mass measurements have their own limitations. For example, they assume the same \gls{IMF} for all galaxies, which is unlikely to be realistic. However, only working with galaxies that have consistent stellar mass estimates using both techniques provides some additional certainty to the reliability of the measurements used in this work. Thus, we apply the following additional consistency threshold: 

\begin{equation}
    \vert \log \left(M_{\rm *, Re, JAM}\right) - \log \left(M_{\rm *, Re, SPS}\right) \vert < 0.5\;.
\end{equation}

This threshold is visualized in the rightmost panel of Figure \ref{fig:consistency}. Applying all these consistency thresholds to our sample further reduces the sample size to \todo{1,362} galaxies.

The \gls{TW} method also comes with its own set of assumptions and limitations. For example, it can only be used on galaxies with intermediate inclinations, regular kinematics, and where the position angle of the bar is not parallel or perpendicular to the position angle of the disc. \citet{geron_2023} looked at the same 10,010 unique galaxies in the \gls{MaNGA} survey, but was only able to measure the bar kinematics for 225 galaxies due to these restrictions. Despite this large drop in sample size, it is worth keeping in mind that \citet{geron_2023} is still the largest catalog of measurements of bar kinematics to date. 

In this work, we further reduce the sample size by only using galaxies with the most reliable measurements by removing galaxies with extremely large uncertainties on the bar pattern speed and $\mathcal{R}$. We exclude all galaxies where the $1\sigma$ uncertainty on the bar pattern speed is greater than 10 km s$^{-1}$ kpc$^{-1}$ or the $1\sigma$ uncertainty on $\mathcal{R}$  is greater than 1. This reduced the sample to \todo{126} galaxies.


Finally, we cross-match the galaxies from \citet{zhu_2023} that pass the quality and consistency thresholds with the remaining galaxies from \citet{geron_2023}, which results in a final sample of \todo{30} unique galaxies. The change in sample size due to all thresholds is summarized in Table \ref{tab:samplesize}.

\begin{figure}
  \centering
    \includegraphics[width=1.0\columnwidth]{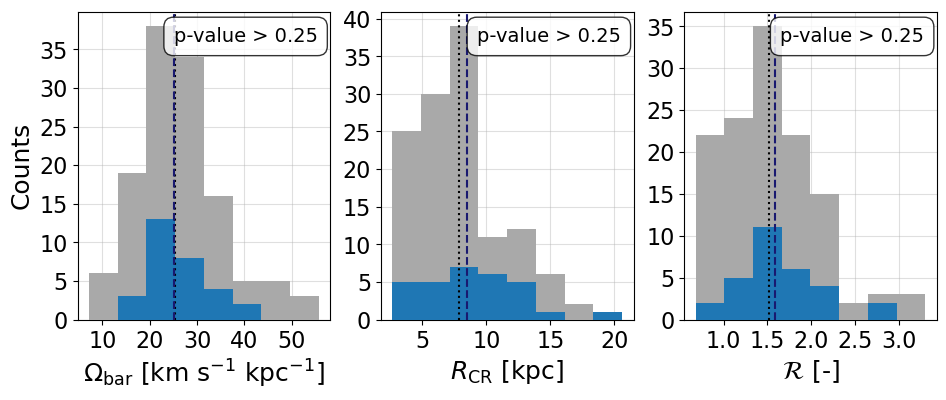}
    \caption{A comparison of the bar pattern speed (left panel), corotation radius (middle panel), and $\mathcal{R}$ (right panel) for the final sample used in this paper (blue) and all the \gls{TW} galaxies that pass the quality thresholds (gray) The median values of both samples is denoted by the dashed and dotted lines, respectively. The p-value of an Anderson-Darling test is shown in the top-right corner of each panel, and is $>0.25$ for all comparisons. This suggests that the final sample used in this work is not obviously biased.}
    \label{fig:tw_sample}
\end{figure}

\begin{figure}
  \centering
    \includegraphics[width=1.0\columnwidth]{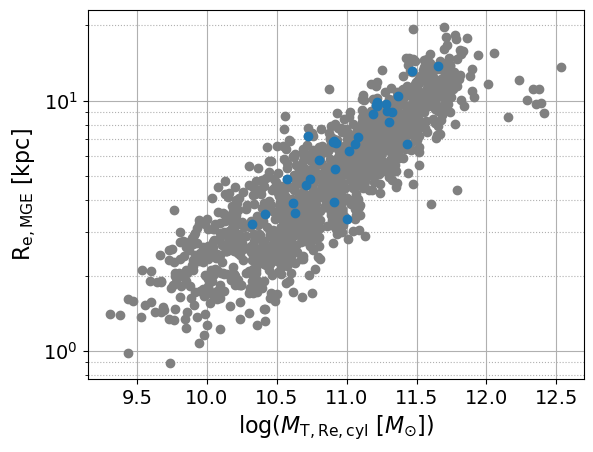}
    \caption{A size ($R_{e}$) - mass ($M_{\rm T, Re}$) diagram for all galaxies in the final sample used in this work (blue) and all \gls{JAM} models that pass all the quality and consistency thresholds (gray). The final sample is not clearly biased towards any particular size or mass.}
    \label{fig:size_mass}
\end{figure}

To make sure the final cross-match of both catalogs does not bias the sample, we compare the distribution of the three bar kinematics parameters of the final sample to the distribution of the \gls{TW} galaxies that pass the quality thresholds in Figure \ref{fig:tw_sample}. The median values of both samples are very similar and the p-value of an Anderson-Darling test is $>$0.25 for all three parameters. It is clear that the distributions of both samples are very similar and we find no evidence that our sample is biased. We also show the final sample on a size-mass diagram in Figure \ref{fig:size_mass} and compare it to the \gls{JAM} models that pass all quality and consistency thresholds. While the mass and radius range of the final sample is more limited than the full range covered by all the \gls{JAM} models, it is also clear that the final sample is not biased towards any particular size or mass.

\begin{deluxetable}{cccc}
\label{tab:samplesize}
\tablecaption{An overview of the change in sample size after applying the multiple thresholds described in Section \ref{sec:sample_selection}. Both the JAM measurements from \citet{zhu_2023} and TW measurements from \citet{geron_2023} start with the full MaNGA galaxy sample. However, we apply different quality and consistency thresholds to both, before combining them to create a final sample of \todo{30} galaxies.}
\tablehead{\colhead{JAM Sample} & \colhead{} & \colhead{TW Sample} & \colhead{}}
\startdata
Full MaNGA Sample & \multicolumn{1}{c|}{10,010} & Full MaNGA Sample & 10,010 \\ 
Quality Thresholds & \multicolumn{1}{c|}{4,867} & TW Thresholds & 255 \\ 
Consistency Thresholds & \multicolumn{1}{c|}{1,362} & Quality Thresholds & 126 \\ 
 \hline
 Final Combined Sample & 30  & &\\ 
\enddata
\end{deluxetable}

\section{Results}
\label{sec:results}

We have a sample of \todo{30} galaxies with reliable bar kinematics measurements and reliable \gls{JAM} models. In Section \ref{sec:tw_vs_jam}, we explore how these two sets of parameters compare against each other. We also study whether the NFW halo parameters are affected by the kinematics of the bar in Section \ref{sec:nfw_params}.


\subsection{Tremaine-Weinberg against Jeans Anisotropic Modeling}
\label{sec:tw_vs_jam}

\begin{figure*}
  \centering
    \includegraphics[width=1.0\textwidth]{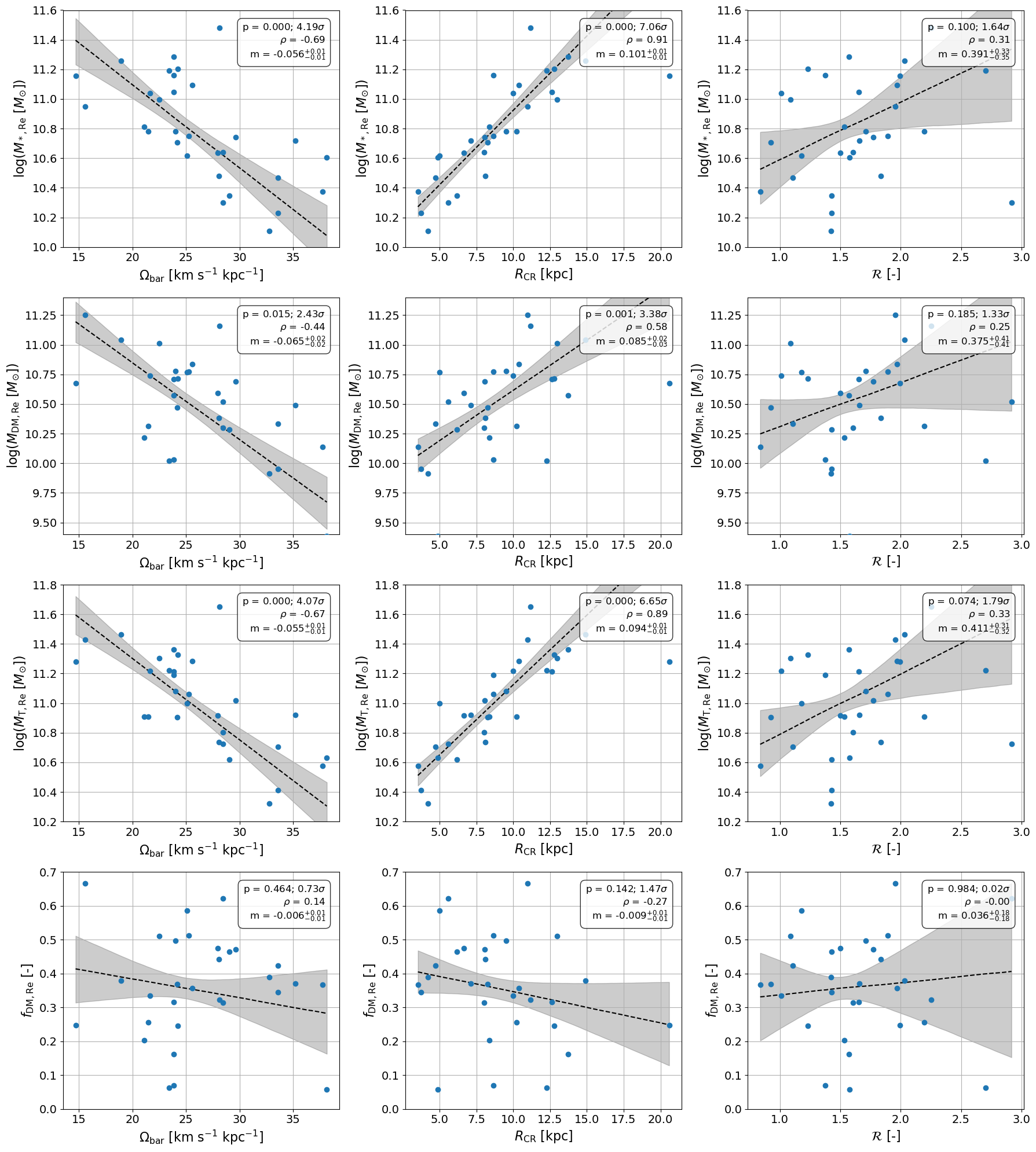}
    \caption{We plot the different bar kinematics parameters ($\Omega_{\rm b}$, left column; $R_{\rm CR}$, middle column; $\mathcal{R}$, right column) against the different JAM parameters ($\log \left(M_{\rm *, Re}\right)$, top row; $\log \left(M_{\rm DM, Re}\right)$, top middle row; $\log \left(M_{\rm T, Re}\right)$, bottom middle row; $f_{\rm DM, Re}$, bottom row). The significance of each correlation was tested with a Spearman test, and its p-value is shown in the top-right corner of each plot. We also fit a line through the data using \texttt{linmix}, with the $1\sigma$ contours shown in grey. We find that $\Omega_{\rm b}$ and $R_{\rm CR}$ are correlated with $\log \left(M_{\rm *, Re}\right)$, $\log \left(M_{\rm DM, Re}\right)$, and $\log \left(M_{\rm T, Re}\right)$. However, we do not find any significant correlation between $f_{\rm DM, Re}$ and any of the bar kinematics parameters. Similarly, we do not find any significant correlation between $\mathcal{R}$ and any of the \gls{JAM} parameters.}
    \label{fig:tw_vs_jam}
\end{figure*}

The bar kinematics parameters are plotted against the different masses measured with the \gls{JAM} dynamical models in Figure \ref{fig:tw_vs_jam}. We perform a Spearman correlation test on each combination to assess whether the two parameters are correlated. We also fit a line through the data using \texttt{linmix}\footnote{\url{https://linmix.readthedocs.io/en/latest/index.html}} \citep{kelly_2007}, a Bayesian MCMC-based method that is able to account for heteroscedastic uncertainties
in both the $x$ and $y$ axes.


As shown in the leftmost column of Figure \ref{fig:tw_vs_jam}, the bar pattern speed is significantly anti-correlated with the stellar mass and total dynamical mass (\todo{p-value $<$ 0.0001; $>4.0\sigma$}), suggesting that bars with the lowest values of $\Omega_{\rm b}$ reside in the highest mass galaxies (in terms of stellar mass and total dynamical mass). Interestingly, while the best-fit line between the bar pattern speed and the dark matter mass is negative (\todo{m = -0.07$^{+0.02}_{-0.02}$}), the Spearman test reveals that it falls just short of the default $3\sigma$ threshold (p-value = 0.015; $2.43\sigma$). These contradicting results suggest that more data is needed to clarify the significance of this trend. The bar pattern speed is not correlated with the dark matter fraction. 

The trends between the corotation radius and the different masses are shown in the middle column of Figure \ref{fig:tw_vs_jam}. The corotation radius shows a statistically significant correlation with the stellar mass, dark matter mass, and total dynamical mass (\todo{$>3\sigma$} in all cases). This implies that higher mass galaxies have larger corotation radii. However, no correlation was found with the dark matter fraction in this case either.

We do not find any significant correlation between $\mathcal{R}$ and any of the \gls{JAM} parameters, as shown in the rightmost column of Figure \ref{fig:tw_vs_jam}. The slope of the best-fit line is still positive for all the different masses, but the $1\sigma$ uncertainties on the fit only barely exclude a flat (m = 0) line.

\subsection{NFW Parameters}
\label{sec:nfw_params}

\begin{figure*}
  \centering
    \includegraphics[width=1.0\textwidth]{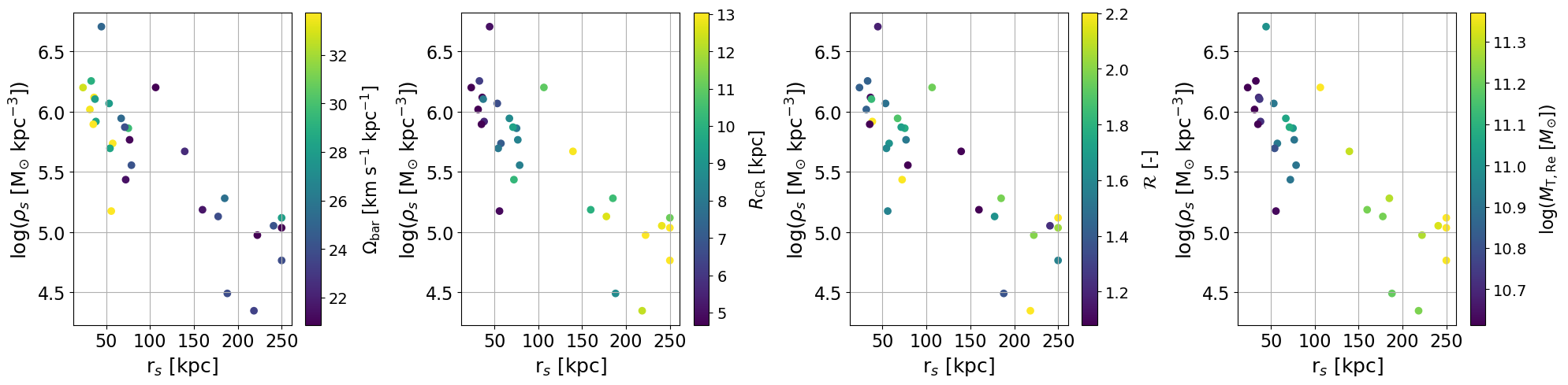}
    \caption{We show the NFW halo parameters $r_s$ and $\log \left( \rho_s \right)$ against each other. We additionally color each point based on its bar pattern speed (leftmost panel), corotation radius (middle left panel), $\mathcal{R}$ (middle right panel), and total dynamical mass (rightmost panel). The galaxies with the highest values of $\Omega_{\rm b}$ tend to have high values of $\log \left( \rho_s \right)$ and low values of $r_s$, while this is inverted for the corotation radius. No discernible trend is found for $\mathcal{R}$.}
    \label{fig:nfw_params}
\end{figure*}

As discussed in Section \ref{sec:jam}, the dark matter halo is parameterized with a standard NFW dark matter halo profile \citep{navarro_1996} with two parameters: the characteristic density ($\rho_s$) and the characteristic radius ($r_s$). Figure \ref{fig:nfw_params} shows how these two parameters are affected by the three bar kinematics parameters (bar pattern speed, corotation radius, and $\mathcal{R}$) and the total dynamical mass obtained from the \gls{JAM} models. Interestingly, we see in the leftmost panel that the galaxies with the highest values of $\Omega_{\rm b}$ (i.e., the galaxies with bars that rotate the fastest) tend to have high values of $\log \left( \rho_s \right)$ and low values of $r_s$. Conversely, galaxies with low values of $\Omega_{\rm b}$ (i.e., the galaxies with bars that rotate the slowest) have low values for $\log \left( \rho_s \right)$ and high values for $r_s$. 

We find a similar, but inverted trend for the corotation radius in the second panel of Figure \ref{fig:nfw_params}: galaxies with high corotation radii tend to have NFW haloes with low values for $\log \left( \rho_s \right)$ and high values for $r_s$. Interestingly, no clear trends are found for $\mathcal{R}$ in the third panel of Figure \ref{fig:nfw_params}.

\section{Discussion}
\label{sec:discussion}

\subsection{Evidence for Bar Slowdown with the Bar Pattern Speed}
\label{sec:disc_omega}

Many simulations predict that bars slow down due to the exchange of angular momentum between the bar and the outer disc and dark matter halo. Thus, galaxies with a more massive disc or dark matter halo can slow down the bar more efficiently \citep{lynden_bell_1972,sellwood_1981,athanassoula_2003,sellwood_2008, athanassoula_2013, semczuk_2024}. 

This is in agreement with the results presented in this work. We find that the bar pattern speed is correlated with stellar mass and total dynamical mass, as shown in the left column of Figure \ref{fig:tw_vs_jam}. Additionally, we find a tentative correlation between the bar pattern speed and the dark matter mass, though more work is needed to confirm this latter result. Either way, we find that bars tend to be slower (in terms of $\Omega_{\rm b}$) in more massive galaxies.

Interestingly, the correlation between the bar pattern speed and stellar mass is more significant than the correlation between the bar pattern speed and dark matter mass. This might be because the dark matter mass measurements are intrinsically more uncertain than the stellar mass measurements (see Sections \ref{sec:jam} and \ref{sec:sample_selection}), or perhaps the stellar component is more efficient at slowing down the bar. Previous work has specifically highlighted the role of the outer disc \citep{sellwood_1981,athanassoula_2003} and bulge \citep{kataria_2019, li_2024, mcclure_2025} in slowing down the bar.

There is some noise in these correlations. However, this is expected, as we only observe the bar pattern speed at one moment in time. We do not have any information on the initial bar pattern speed, the slowdown rate, or the age of the bar, which all contribute to the noise in the correlation. For example, when we see a slow bar, we do not know whether it was slowed down or simply formed slow. Additionally, since we also do not know when the bar was formed, we do not have constraints on how long the bar has been slowing down. This makes comparison between different galaxies difficult, a concern that was also raised by \citet{guo_2019}. This degeneracy can be broken by additionally measuring the age of the bar using techniques such as the one presented in \citet{de_sa_freitas_2025}.


We do not find any high-mass galaxies with fast (high $\Omega_{\rm b}$) bars. This is expected, as the massive disc and halo would quickly slow down the bar. However, we also do not find a population of low-mass galaxies with slow (low $\Omega_{\rm b}$) bars. This is more puzzling, as one could expect some bars to be formed with low pattern speeds. In fact, \citet{semczuk_2024} find that high-mass galaxies form bars with high initial pattern speeds that quickly slow down, in agreement with our observations. In contrast, they find that low-mass galaxies form bars with low pattern speeds that do not slow down. This population of barred galaxies is not found in this work. 


We cannot exclude the possibility that there are unknown systematics in the data that bias both the \gls{TW} method and \gls{JAM} models, which might affect our conclusions. While this scenario seems unlikely, as we only work with the most robust measurements (see Section \ref{sec:sample_selection}), it will be useful to verify the findings in this work using different techniques to probe the bar kinematics and stellar, dark matter, and total dynamical mass, such as the barred Schwarzschild method \citep{tahmasebzadeh_2022, jin_2025b, jin_2025}.

Finally, we note that this result is in disagreement with the results presented in \citet{geron_2023}, who found no correlation between the bar pattern speed and stellar mass. This discrepancy likely stems from differences in sample selection. As noted in Section \ref{sec:sample_selection}, the sample of bar kinematics measurements used in this work is effectively a subset of the sample used in \citet{geron_2023}. However, their full sample includes targets with very large uncertainties associated with the bar pattern speed measurements, which diluted the correlation. The results presented in this work are more robust, as we only use targets with high-quality measurements and properly incorporate the measurement uncertainties in the analysis.

\subsection{Corotation Radius and the Tully-Fisher Relation}
\label{sec:disc_rcr}

As shown in the middle column of Figure \ref{fig:tw_vs_jam}, we find statistically significant correlations between the corotation radius and the stellar mass, dark matter mass, and the total dynamical mass. Thus, barred galaxies with larger corotation radii tend to be more massive. As a bar can grow up to its corotation radius, this finding is in agreement with other observational studies that find that longer bars are typically found in more massive galaxies (e.g., \citealp{gadotti_2011,erwin_2019}). 

These correlations can be explained with the (baryonic) Tully-Fisher relation \citep{tully_1977,mcgaugh_2000}. As the corotation is often well approximated by $R_{\rm CR} \approx V_{\rm c} / \Omega_{\rm b}$, and the Tully-Fisher relation effectively correlates the rotation velocity with the total baryonic mass of a galaxy, we expect to see a correlation between the corotation radius and the mass of the galaxy. Any additional deviations from this are likely caused by differences in the bar pattern speed.


\subsection{The Limitations of $\mathcal{R}$}
\label{sec:disc_r}

While we find statistically significant correlations with the bar pattern speed, we do not find any trends between $\mathcal{R}$ and any of the masses tested in this work (stellar mass, dark matter mass, total dynamical mass, and the dark matter fraction). While the best-fit slope is positive, the $1\sigma$ do not (or barely) exclude a flat (m=0) line. This is in agreement with \citet{guo_2019}, who also found no correlation between $\mathcal{R}$ and the dark matter fraction. 

However, while we find no evidence for the slowing down of bars as quantified by $\mathcal{R}$, we argue that this is possibly due to inherent difficulties associated with measuring this parameter. It depends on estimates of the bar pattern speed, the rotation curve of the galaxy, and the bar radius, which can all introduce biases and uncertainties. Additionally, the bar is expected to slow down over time, which increases its corotation radius, while at the same time becoming longer. As $\mathcal{R} = R_{\rm CR} / R_{\rm bar}$, both the denominator and numerator are expected to increase, and thus their ratio depends on which parameter increases faster. All of this can result in noisy and unstable measurements. 

$\mathcal{R}$ has also been at the forefront of multiple controversies. For example, many observations find so-called `ultrafast bars', which have $\mathcal{R} < 1$ (e.g., see \citealp{aguerri_2015, guo_2019, garma_oehmichen_2020}). However, these bars are not supposed to exist, as bars cannot grow beyond their corotation radius \citep{contopoulos_1980,contopoulos_1981, athanassoula_1992}. Additionally, simulations predict relatively high values for the distribution of $\mathcal{R}$, while observations tend to find significantly lower values. This discrepancy has been raised as a tension for $\Lambda$CDM \citep{roshan_2021b}. The observations of \citet{geron_2023} find higher overall values of $\mathcal{R}$ compared to previous observational work, reducing both aforementioned problems. However, the tension remains, and they still find that $2\%$ of bars are confidently within the ultrafast regime. \citet{cuomo_2021} suggest another solution to these problems: they find that the bar length obtained from observations is often wrong, due to apparent associations with rings and spiral arms. This changes subsequent measurements of $\mathcal{R}$, as it directly depends on the bar length. In a similar vein, \citet{frankel_2022} noted that bars in simulations are not slower, but instead shorter than observed bars. 


Thus, $\mathcal{R}$ is a notoriously difficult parameter to measure accurately. All these uncertainties can compound and hide any trends, which explains why we see correlations with the bar pattern speed, but not $\mathcal{R}$. \citet{semczuk_2024} also come to a similar conclusion, and give preference to using the bar pattern speed in their work, as opposed to $\mathcal{R}$. 



\subsection{No Correlations with the Dark Matter Fraction}

We do not find any correlations between the dark matter fraction and any of the bar kinematics parameters. However, similar to $\mathcal{R}$, the dark matter fraction is a hard parameter to quantify accurately. Since both the dark matter mass and the total dynamical mass span a wide range of values and are typically expressed on a logarithmic scale, a small change in either can result in a large change in the dark matter fraction. Similarly, a small increase in the uncertainty in either parameter can result in a large uncertainty in the dark matter fraction. This can result in large measurement errors that vary between models, which is clearly visible in the fourth panel of Figure \ref{fig:consistency}.


Additionally, as shown in the first column of Figure \ref{fig:tw_vs_jam} and discussed in Section \ref{sec:disc_omega}, we find a correlation between the bar pattern speed and the total dynamical mass, as well as a tentative correlation with the dark matter mass. Since both the denominator and numerator depend on the bar pattern speed, a trend in their ratio will be more difficult to detect. 


Thus, while we do not find any correlation with the dark matter fraction in this work, we do not exclude the possibility that our current methods are not sensitive enough. Revisiting this problem with a larger sample of galaxies will be necessary to confirm these conclusions. The galaxies used in this work are all nearby extended galaxies observed by the \gls{MaNGA} survey, with a stellar mass range between $10^{10} - 10^{11.5}$M$_{\odot}$. A good strategy for future work would be to supplement this sample with observations over a wider range of galaxy types and stellar (and dark matter) masses, such as dwarf galaxies with potentially large dark matter fractions (e.g., see \citealt{cuomo_2024}). 




\subsection{NFW Model Parameters}
\label{sec:disc_fdm}

Figure \ref{fig:nfw_params} suggests that galaxies with low values of $\Omega_{\rm b}$ (i.e., galaxies with bars that rotate the slowest) have low values for $\log \left( \rho_s \right)$ and high values for $r_s$. In contrast, galaxies with the highest values of $\Omega_{\rm b}$ (i.e., galaxies with bars that rotate the fastest) tend to have high values of $\log \left( \rho_s \right)$ and low values of $r_s$. This initially seems counterintuitive, as this implies that slower bars (in terms of $\Omega_{\rm b}$) have less dark matter in the centers of their galaxies. However, their dark matter halo also extends farther, which results in an overall higher dark matter mass that can slow down the bar more efficiently. Thus, slower bars tend to have a more extended (high $r_s$) NFW dark matter profile with a lower central density (low $\rho_s$). Interestingly, no pattern is found between $\mathcal{R}$ and the NFW halo parameters. 

However, as noted in Section \ref{sec:jam}, it is worth keeping in mind that $r_s$ is not a free parameter in the models used in this work. It is fixed using the well-established mass–concentration relation \citep{dutton_2014}, while $\rho_s$ remains a free parameter. This is shown in the rightmost panel of Figure \ref{fig:nfw_params}, where you can clearly see that the most massive galaxies have higher values of $r_s$, as expected. This explains the trends seen with the corotation radius, which is also tightly correlated with the mass of the galaxy. This suggests that these trends could simply reflect the imposed scaling relations. Future studies can expand on this work and verify the results presented here by creating models that keep both $r_s$ and $\rho_s$ as free parameters.



\section{Conclusion}
\label{sec:conclusion}



Simulations predict that the bar should slow down due to angular momentum exchange with the outer disc and dark matter halo \citep{lynden_bell_1972,sellwood_1981,athanassoula_2003,sellwood_2008,athanassoula_2013, kataria_2019, beane_2023, li_2023, li_2024, semczuk_2024}. This implies that galaxies with a more massive dark matter halo or disc should be able to slow down the bar more efficiently. However, observational evidence for this has been limited (e.g., see \citealp{guo_2019,buttitta_2023, tahmasebzadeh_2024}). We present, for the first time with a larger sample of galaxies (n=30), observational evidence for the slowdown of bars due to angular momentum exchange with its host. We combine bar kinematics measurements made by \citet{geron_2023} using the \gls{TW} method with \gls{JAM} models developed by \citet{zhu_2023}. Our main results are:


\begin{itemize}
    \item We find a statistically significant correlation between the bar pattern speed and the stellar mass (\todo{$4.19\sigma$}), as well as the total dynamical mass (\todo{$4.07\sigma$}). Slower bars (low $\Omega_{\rm b}$) tend to be hosted by more massive galaxies. We also find a tentative correlation with the dark matter mass (\todo{$2.43\sigma$}), though additional work is needed to confirm this. 

    \item We find a positive correlation between the corotation radius and stellar mass (\todo{$7.06\sigma$}), dark matter mass (\todo{$3.38\sigma$}), and total dynamical mass (\todo{$6.65\sigma$}). More massive galaxies tend to host bars that have larger corotation radii. These correlations are expected through the (baryonic) Tully-Fisher relation.

    \item Interestingly, we do not find any correlations that involve $\mathcal{R}$ or the dark matter fraction. This is possibly due to the multiple challenges associated with measuring these parameters accurately. 
    
    \item Slower bars (low $\Omega_{\rm b}$) have more extended (high $r_s$) NFW dark matter profiles with lower central densities (low $\rho_s$), compared to faster (high $\Omega_{\rm b}$) bars. We did not find a correlation between $\mathcal{R}$ and the NFW halo parameters.

\end{itemize}

These results are the first step towards studying how kinematics of bars depend on host galaxy properties observationally on a larger scale. However, more work will be necessary to confirm some of the results presented in this work. For example, we can extend the sample to include galaxies with different stellar masses, such as dwarf galaxies, or measure the bar kinematics and dark matter mass with other techniques (e.g., \citealp{tahmasebzadeh_2022}). We also did not address how gas affects bar kinematics in this work, which could influence some of the conclusions. For example, \citet{beane_2023} found that adding gas to their simulations stabilizes the bar pattern speed and results in bars that do not slow down. We will address this in future work by using observations from the \gls{ALFALFA} survey \citep{giovanelli_2005}. 


\begin{acknowledgments}
TG is a Canadian Rubin Fellow at the Dunlap Institute. The Dunlap Institute is funded through an endowment established by the David Dunlap family and the University of Toronto.

\end{acknowledgments}

\software{\texttt{astropy} \citep{astropy_2013, astropy_2018, astropy_2022}, \texttt{JamPy} \citep{cappellari_2020}, \texttt{linmix}\footnote{\url{https://linmix.readthedocs.io/en/latest/index.html}}, \texttt{Matplotlib} \citep{matplotlib_2007}, \texttt{MgeFit} \citep{cappellari_2002}, \texttt{numpy} \citep{numpy_2020}, \texttt{scipy} \citep{scipy_2020}, \texttt{Tremaine-Weinberg} \citep{geron_tw_2023}}







\bibliography{bibtex}{}
\bibliographystyle{aasjournal}



\end{document}
